\documentclass[
aps,
prd,
reprint,
10pt,
superscriptaddress,
nofootinbib,
longbibliography
]{revtex4-2}

\usepackage{amsmath,amssymb,amsthm,mathtools}
\usepackage{bm}
\usepackage{physics}
\usepackage{bbm}
\usepackage{xcolor}
\providecommand{\texorpdfstring}[2]{#1}

\theoremstyle{plain}
\newtheorem{proposition}{Proposition}
\theoremstyle{remark}
\newtheorem{remark}{Remark}
\theoremstyle{definition}

\begin{document}

\title{Regularization from Superpositions of Time Evolutions}

\author{Yakir Aharonov}
\affiliation{Schmid College of Science and Technology, Chapman University, Orange, California 92866, USA}
\affiliation{Institute for Quantum Studies, Chapman University, Orange, California 92866, USA}
\affiliation{School of Physics and Astronomy, Tel Aviv University, Tel Aviv 6997801, Israel}

\author{Eliahu Cohen}\thanks{Corresponding author: eliahu.cohen@biu.ac.il}
\affiliation{Faculty of Engineering and the Institute of Nanotechnology and Advanced Materials, Bar-Ilan University, 5290002 Ramat Gan, Israel}
\affiliation{Institute for Quantum Studies, Chapman University, Orange, California 92866, USA}

\author{Tomer Shushi}
\affiliation{Center for Quantum Science and Technology, Department of Business Administration, Guilford Glazer Faculty of Business and Management, Ben-Gurion University of the Negev, Beer-Sheva, Israel}

\begin{abstract}
Short-time approximations and path integrals may be dominated by high-energy or large-field contributions, especially for singular interactions, motivating regulators that suppress such contributions while remaining removable. We show that smooth regulators of this kind can arise as conditional maps generated by interference in coherently controlled, postselected superpositions of time evolutions. In quantum mechanics, a Gaussian superposition of short-time translations produces a Gaussian energy filter that renders time-sliced kernels well behaved for singular potentials while recovering the target unitary dynamics when the regulator is removed and in the continuum limit. We also discuss a related momentum-space regulator obtained by coherently smearing a kinetic prefactor. In scalar quantum field theory, local Gaussian smearing of the quartic coupling induces a positive local $\phi^8$ term in the Euclidean action, providing a symmetry-compatible large-field stabilizer. We derive short-time error bounds, analyze multi-step success probabilities, and discuss how the construction can be used as a removable regulator together with conventional ultraviolet regularization.
\end{abstract}

\maketitle

\section{Introduction}

Unitary time evolution generated by a self-adjoint Hamiltonian is the cornerstone of quantum theory.
In many practical calculations, however, one works with \emph{short-time} propagators (for time slicing,
semi-classical approximations, or path-integral constructions) or with functional integrals over field
configurations. In these intermediate steps, singular interactions and extreme configurations can make
standard short-time approximations or weights ill-behaved. This motivates regulators that suppress such
extreme contributions, yet remain removable so that the target dynamics is recovered. The proposed novelty lies in the \emph{operational} origin for such regulators: the suppressing factors arise
as the \emph{conditional} map produced by interference in a coherently controlled, postselected
superposition of evolutions, rather than being inserted ad hoc.

Compared with standard regularization schemes (momentum/energy cutoffs, lattice regularization,
Pauli--Villars fields, dimensional regularization, point splitting, or higher-derivative/heat-kernel
smearing) our construction is not only a convenient mathematical insertion: the suppressing factor is the
Kraus operator of a physically implementable coherent-control and postselection protocol. In the time-translation
case, a finite width in the superposed short steps can be viewed as limited temporal resolution of a quantum
control/clock degree of freedom: large-energy components acquire rapidly varying phases across the superposition
and therefore interfere destructively upon postselection, producing the Gaussian suppression
$e^{-(\sigma^2\Delta t^2/2)H^2}$. In the QFT setting, local smearing of $\lambda(x)$ plays a closely related
role, mimicking short-distance or environmental fluctuations of effective couplings and yielding a local
higher-dimensional stabilizer term. Throughout the text we adopt the conservative viewpoint that $\sigma$ is a removable
regulator parameter. Any interpretation as a fundamental high-energy modification would require separate
consistency checks.

The construction is related to several established ideas,
including linear combinations of unitaries in quantum
algorithms \cite{ChildsWiebe2012,Berry2015,LowChuang2017},
conditional nonunitary dynamics in continuous-measurement
theory \cite{BarchielliBelavkin1991,JacobsSteck2006,
WisemanMilburn2009}, spectral and heat-kernel
regularization \cite{Seeley1967,Vassilevich2003},
non-Hermitian damping and absorbing potentials
\cite{Moiseyev2011,RissMeyer1995,Muga2004}, and
large-field stabilization in constructive Euclidean field
theory \cite{Simon1974,GlimmJaffe1987,Rivasseau1991}.
We do not claim novelty for damping factors or stabilizing
terms themselves. Rather, the point of the present work is
their operational origin: the regulator arises as a
trace-decreasing Kraus map generated by coherent control
and postselection and is treated conservatively as
removable. A more detailed comparison with the
literature is given in Sec.~\ref{sec:relation-existing-methods}.

We propose an \emph{interference-based} regularization mechanism built from coherent superpositions of
time evolutions. The operational starting point is the ``superposition of time translations'' protocol of
Aharonov, Anandan, Popescu and Vaidman (AAPV) \cite{AAPV1990}, which introduced the idea of superposing
\emph{time-evolution operators} (rather than states). In its simplest form, one engineers complex coefficients
$\{a_j\}$ (rescaled so that $\sum_j a_j=1$, absorbing any overall scalar into the heralding amplitude)
such that, on a chosen family of input states, a superposition of evolutions behaves effectively like a
single evolution,
\[
\sum_j a_j\,U_j\ket{\psi}\;\approx\;U'\ket{\psi},
\qquad \sum_j a_j=1.
\]
Operationally, the superposition is obtained by correlating the evolutions $U_j$ with an ancilla basis and
postselecting the ancilla, so that the conditioned system map is a single Kraus operator proportional to
$\sum_j a_j U_j$, with coefficients set by the pre- and postselected ancilla amplitudes.
AAPV emphasized that suitable choices of these coefficients can yield effective evolutions far outside the range of the
constituents (including amplification of weak forces and effective time translation by durations that can even be
negative).

In the present work we use the same conditioned-interference mechanism in a complementary regime:
we choose superpositions narrowly peaked around the desired short-time step, and the leading finite-width
deviation becomes a controlled, \emph{removable} damping factor that we use for regularization. We develop this
mechanism in two settings.

\noindent\textit{Quantum mechanics.}
We study a Gaussian superposition of short-time steps, characterized by a width parameter $\sigma$\footnote{We use the same symbol $\sigma$ in the QM and QFT parts. In QM it is the (dimensionless) width of the time-step distribution; in QFT it controls the strength of the induced local operator after averaging over a coupling.}.
The resulting postselected step is
\[
V_{\sigma,\Delta t}=e^{-iH\Delta t}\,e^{-\frac12\sigma^2\Delta t^2H^2},
\]
which damps high-energy components on the scale $|E|\gtrsim 1/(\sigma\Delta t)$ while reducing to the
unitary step as $\sigma\to 0$. We use this map as a \emph{short-time regulator} for kernels appearing in
time-sliced path-integral approximations in the presence of singular potentials. A closely related QM variant,
developed in Sec.~3, replaces the time smearing by a coherent slice-by-slice smearing of a kinetic prefactor
(or mass fluctuation), producing a smooth $p^4$ suppressor of large momenta.

\noindent\textit{Quantum field theory.}
We show that an analogous \emph{local} Gaussian averaging of the quartic coupling $\lambda(x)$ in scalar
theory induces a positive $(\sigma^2/2)\phi^8(x)$ term in the Euclidean action. This irrelevant operator
preserves the symmetries of the scalar model and stabilizes large-field tails of the measure. As in the
QM case, the construction is used as a removable regulator: renormalization is carried out at fixed
$\sigma$, and the limit $\sigma\to 0$ recovers the target theory.

The technical core of the manuscript is an explicit short-time expansion and error bound for postselected
linear combinations of evolutions, together with a discussion of the heralding (postselection)
probability and its behavior under multiple steps. These results quantify when the interference-induced
filter is negligible and when it becomes significant.

The manuscript is organized as follows. Section~2 introduces the general postselected linear-combination
map, derives a short-time expansion, and specializes to Gaussian time-smearing. Section~3 discusses a
closely related variant in which one coherently smears a kinetic prefactor (equivalently, a mass
fluctuation) in each short-time slice, yielding a smooth momentum suppressor. Section~4 applies the
resulting energy-filtered short-time kernel to representative singular quantum-mechanical examples.
Section~5 shows how a local Gaussian smearing of couplings in scalar QFT induces a stabilizing $\phi^8$
interaction. Section~6 places the construction in relation to existing
regularization, measurement, and non-Hermitian approaches.
Section~7 summarizes the the work and outlines
possible extensions.

\section{Superpositions of time evolutions: short-time control and Gaussian filters}

We begin with a general controlled-evolution/postselection construction that produces a (generally non-unitary)
Kraus operator given by a coherent linear combination of unitary evolutions, and then specialize to the
Gaussian time-smearing used in later sections.

Let $\{H_j\}_{j=1}^m$ be self-adjoint operators (bounded or defined on a common dense domain) and let
$\{a_j\}$ be complex coefficients with $\sum_j a_j = 1$. For a small time step $\Delta t$ define
\begin{equation}\label{eq:Umix-def}
U_{\text{mix}}(\Delta t) := \sum_{j=1}^m a_j\, e^{-i H_j \Delta t}, \qquad
\bar H := \sum_{j=1}^m a_j H_j.
\end{equation}

In an AAPV implementation, $U_{\text{mix}}(\Delta t)$ is (up to an overall complex scalar) the Kraus
operator associated with a successful postselection on a control ancilla. Let $A$ denote the ancilla and $S$
the system. Prepare
$\ket{\chi}_A=\sum_j c_j\ket{j}_A$ and an input state $\ket{\psi}_S$, apply the controlled evolution
\[
U_{A\to S}(\Delta t):=\sum_j \ket{j}\!\bra{j}_A\otimes e^{-iH_j\Delta t},
\]
and postselect the ancilla onto $\ket{\phi}_A=\sum_j \phi_j\ket{j}_A$. Then
\begin{align*}
\ket{\chi}_A\otimes\ket{\psi}_S
&\xrightarrow{U_{A\to S}(\Delta t)}
\sum_j c_j\ket{j}_A\otimes e^{-iH_j\Delta t}\ket{\psi}_S,\\
{}_A\!\bra{\phi}(\cdot):\ \ket{\psi}_S
&\mapsto
\Big(\sum_j \phi_j^\ast c_j\,e^{-iH_j\Delta t}\Big)\ket{\psi}_S,\\
&=U_{\text{mix}}(\Delta t)\ket{\psi}_S.
\end{align*}
so that $a_j=\phi_j^\ast c_j$ in \eqref{eq:Umix-def}. In general $\sum_j a_j=\braket{\phi}{\chi}$, so our
convention $\sum_j a_j=1$ is obtained by rescaling all coefficients by the (nonzero) overlap $\braket{\phi}{\chi}$,
which only changes the overall heralding amplitude (success probability) and not the normalized postselected state.
Any overall complex scalar can be absorbed into the
heralding amplitude and does not affect the normalized postselected state. (For algorithmic uses of related linear-combination-of-unitaries
constructions see \cite{ChildsWiebe2012,Berry2015,LowChuang2017}.)

Throughout the manuscript we focus on the (unnormalized) operator-level map because it is exactly where the regularizing
filter appears.
Equivalently, on density matrices the conditioned update is the completely positive,
trace-nonincreasing map $\rho\mapsto U_{\text{mix}}(\Delta t)\,\rho\,U_{\text{mix}}^\dagger(\Delta t)$,
with success probability $\mathrm{Tr}[U_{\text{mix}}^\dagger(\Delta t)U_{\text{mix}}(\Delta t)\rho]$.

\begin{proposition}[Short-time equivalence and error bound]\label{prop:shorttime}
Assume the $H_j$ are bounded and let $M:=\max_j \|H_j\|$. If $\Delta t\,M\ll 1$, then
\begin{equation}
\begin{split}
\big\|U_{\text{mix}}(\Delta t) - e^{-i \bar H \Delta t}\big\|
&\le \frac{\Delta t^2}{2}\,\Big\|\sum_j a_j H_j^2 - \bar H^{\,2}\Big\| \\
&\quad + O\!\big(\Delta t^3\,M^3\big).
\end{split}
\end{equation}
where the implicit constant depends only on $\sum_j |a_j|$.
If all $H_j$ commute and the weights satisfy $a_j\in\mathbb R$ with $a_j\ge 0$ (so that $\{a_j\}$ is a
probability distribution), the leading deviation is governed by the $a$-weighted operator variance
${\rm Var}_a(H):=\sum_j a_j H_j^2 - \bar H^{\,2}$ (a positive semidefinite operator under these assumptions).
\end{proposition}
\noindent\emph{Comment.} If some $H_j$ are unbounded, one can formulate the same second-order expansion on
vectors in the common domain of the $H_j^2$ (or, in the time-rescaling case $H_j=u_jH$, on $\mathrm{Dom}(H^2)$).
Also note that if some $a_j$ are complex then $\bar H=\sum_j a_jH_j$ need not be self-adjoint, so
$e^{-i\bar H\Delta t}$ is not unitary and non-unitary effects can appear already at order $\Delta t$ through
the anti-Hermitian part of $\bar H$. In the Gaussian time-smearing construction below the weights are real and
$\bar H=H$.

\noindent This proposition is proved in Appendix A. Thus a narrow superposition of evolutions acts like a single
unitary to first order in $\Delta t$.\footnote{Compare also with product-formula/Trotter expansions \cite{Trotter1959}.}
In the particular case used for our regulator, $H_j=u_j H$ (time rescalings of the same Hamiltonian), the
deviation is governed by moments of the scalar $u$, leading directly to the Gaussian filter
\eqref{eq:gaussian-filter}.

\subsection{Gaussian superposition of time-translations}

A convenient family is a Gaussian superposition over time scales:
\begin{equation}
\int_{\mathbb{R}} du\, g_\sigma(u)\, e^{-i u H \Delta t},\qquad
g_\sigma(u) = \frac{1}{\sqrt{2\pi}\sigma}\, e^{-\frac{(u-1)^2}{2\sigma^2}}.
\end{equation}
By Gaussian integration,
\begin{equation}
\int du\, g_\sigma(u)\, e^{-i u H \Delta t}
\;=\; e^{-i H \Delta t}\, e^{-\frac{1}{2}\sigma^2 \Delta t^2 H^2}.
\label{eq:gaussian-filter}
\end{equation}

Denote the postselected Kraus operator in \eqref{eq:gaussian-filter} by
\begin{equation}
V_{\sigma,\Delta t}:=e^{-iH\Delta t}\,e^{-\frac{1}{2}\sigma^2 \Delta t^2 H^2}.
\end{equation}

\begin{remark}[Unconditioned channel (for comparison)]
If the same controlled evolution is applied but one does not condition on a particular postselection outcome
(equivalently, the ancilla is discarded), the reduced system state undergoes the incoherent time-smearing channel
\[
\Phi_{\sigma,\Delta t}(\rho)=\int_{\mathbb R}du\,g_\sigma(u)\;e^{-iuH\Delta t}\,\rho\,e^{+iuH\Delta t},
\]
which corresponds to unitary evolution accompanied by energy dephasing. The regularization mechanism studied
in this work is the \emph{postselected} filter $V_{\sigma,\Delta t}$.
\end{remark}

Because the postselected step is implemented by a single Kraus operator, its success probability has a simple form:
\begin{equation}
\begin{split}
V_{\sigma,\Delta t}^\dagger V_{\sigma,\Delta t}
&= e^{-\sigma^2\Delta t^2 H^2},\\
p(\psi)
&=\|V_{\sigma,\Delta t}\ket{\psi}\|^2
=\mel{\psi}{e^{-\sigma^2\Delta t^2 H^2}}{\psi}.
\end{split}
\end{equation}
If one demands successful postselection at each of $N=t/\Delta t$ steps, the unnormalized state is
$V_{\sigma,\Delta t}^N\ket{\psi}$ and the total success probability is
\begin{equation}
\begin{split}
p_N(\psi)
&=\big\|V_{\sigma,\Delta t}^N\ket{\psi}\big\|^2 \\
&=\mel{\psi}{\big(V_{\sigma,\Delta t}^\dagger V_{\sigma,\Delta t}\big)^N}{\psi} \\
&=\mel{\psi}{e^{-\sigma^2 t\,\Delta t\, H^2}}{\psi}.
\end{split}
\end{equation}

\paragraph{State-dependent scales.}
When $H$ is unbounded, it is convenient to quantify accuracy and success probabilities on the spectral support
actually occupied by the input state. Fix $\epsilon\ll 1$ and choose $E_\ast$ such that the out-of-window spectral
weight satisfies
\[
\mel{\psi}{\chi_{\mathbb R\setminus[-E_\ast,E_\ast]}(H)}{\psi}\le \epsilon.
\]
On this window, a sufficient regime for near-unitary behavior in a single short step is
\[
\Delta t\,E_\ast \ll 1,
\qquad
\sigma\,\Delta t\,E_\ast \ll 1,
\]
in which case $p(\psi)\approx 1$ for such low-energy inputs. Conversely, a regime that strongly suppresses
$|E|\gtrsim 1/(\sigma\Delta t)$ in a \emph{single} slice typically makes $p_N$ exponentially small in
$N=t/\Delta t$, which is why we treat $V_{\sigma,\Delta t}$ primarily as a short-time regularizing map rather
than as a claim of an efficient repeated postselection scheme.

In the energy basis the filter multiplies each eigen-amplitude by $\exp\{-\frac{1}{2}\sigma^2 \Delta t^2 E^2\}$,
so high-energy components at short times are suppressed.

\begin{proposition}[Many-step limit and effective generator]\label{prop:continuum}
Let $N=t/\Delta t$. For fixed $\sigma$ one has
\begin{equation}
\begin{split}
\big(e^{-iH\Delta t}\, e^{-\frac{1}{2}\sigma^2 \Delta t^2 H^2}\big)^N
&= e^{-iHt}\, e^{-\frac{1}{2}\sigma^2 t\,\Delta t\, H^2} \\
&\xrightarrow[\Delta t\to 0]{\text{strongly}} e^{-iHt}.
\end{split}
\end{equation}
Under the diffusive scaling $\sigma^2=\kappa/\Delta t$ with $\kappa>0$ held fixed,
\begin{equation}
\big(e^{-iH\Delta t}\, e^{-\frac{1}{2}\sigma^2 \Delta t^2 H^2}\big)^N
\xrightarrow[\Delta t\to 0]{} \exp\!\Big(-iHt-\frac{\kappa t}{2}H^2\Big).
\end{equation}
This proposition is proved in Appendix B. Thus, depending on how $\sigma$ scales with $\Delta t$, the continuum
limit recovers the target unitary or yields a controlled non-Hermitian correction that damps large-$E$ sectors.
\end{proposition}

\begin{remark}[Connection to energy monitoring]
The diffusive scaling limit is formally identical to evolution with an imaginary potential proportional to $H^2$,
and it is closely related to standard continuous-measurement/monitoring models in which energy eigencomponents
are selectively damped \cite{BarchielliBelavkin1991,JacobsSteck2006,WisemanMilburn2009,
BarchielliGregoratti2009}. We mention it mainly as a clean way to keep a finite interference-induced damping in
continuous time; the main focus of the work is the fixed-$\sigma$ short-time regulator.
\end{remark}

\begin{remark}[Fields and short-time propagators]
The same short-time expansion applies to time-sliced propagators in the path integral. When the
$H_j$ commute (as in modewise decompositions of free theories), the leading deviation is governed
by the $a$-weighted variance operator ${\rm Var}_a(H)=\sum_j a_j H_j^2-\bar H^{\,2}$ and can be
evaluated mode-by-mode. No Baker--Campbell--\allowbreak Hausdorff commutator corrections arise because
$U_{\rm mix}$ is a \emph{linear combination} rather than a product of exponentials. For unbounded
field Hamiltonians the same statements hold on appropriate common domains (or with an implicit UV
regulator).
\end{remark}

\section{Kinetic-term regularization from coherent mass fluctuations}
\label{sec:mass-fluct}

Section~2 obtained a universal Gaussian energy filter by coherently smearing the short-time
translation parameter. A closely related alternative is to smear, instead, a parameter \emph{inside}
the Hamiltonian---here a prefactor of the kinetic term---and to do so at each short-time slice
$\Delta t$ in a time-sliced/Trotterized description. Physically, this produces a smooth suppression
of large canonical momenta in the \emph{short-time kernel}, complementing the energy filtering
discussed earlier.

The first two subsections below develop the short-time mass-smearing regulator itself and its accumulation over many
slices. The last two subsections provide an interpretational complement: they show how successful postselection can be
viewed as a momentum-selection dynamics and, in a suitable regime, as the emergence of a classical-like path in the
corresponding Feynman sum over histories.

Let
\[
T:=\frac{p^2}{2m},\qquad H_0:=T+V(\hat x),
\]
with reference mass $m>0$.  We introduce a dimensionless kinetic prefactor
\begin{equation}
H(u):=u\,T+V(\hat x),
\qquad
u=1+\frac{\delta m}{m}.
\label{eq:Hu-def}
\end{equation}

To avoid confusion with the time-smearing width in Sec.~2, we denote by $\sigma_m$ the width of the kinetic
prefactor distribution and take a Gaussian prior centered at $u=1$,
\[
g_{\sigma_m}(u)=\frac{1}{\sqrt{2\pi}\sigma_m}\exp\!\Big(-\frac{(u-1)^2}{2\sigma_m^2}\Big).
\]

\subsection{Slice-by-slice smearing and the induced \texorpdfstring{$p^4$}{p4} suppressor}
\label{subsec:mass-slicewise}

Each short-time
factor is replaced by a coherent superposition over $u$ (implemented via a control ancilla and postselection,
as in Sec.~2). The corresponding postselected single-slice Kraus operator is
\begin{equation}
\begin{split}
V^{(T)}_{\sigma_m,\Delta t}
&:=\int_{\mathbb R}du\,g_{\sigma_m}(u)\;e^{-iH(u)\Delta t} \\
&=\int du\,g_{\sigma_m}(u)\;e^{-i(uT+V)\Delta t}.
\end{split}
\label{eq:Vslicewise-def}
\end{equation}
If this postselection succeeds at each of $N=t/\Delta t$ slices, the unnormalized conditioned evolution is
$\big(V^{(T)}_{\sigma_m,\Delta t}\big)^N$.

\paragraph{Free particle.} For $V=0$, all functions of $T$ commute and the Gaussian integral has the following closed-form:
\begin{equation}
\begin{split}
V^{(T)}_{\sigma_m,\Delta t}
&=\int du\,g_{\sigma_m}(u)\;e^{-iuT\Delta t} \\
&= e^{-iT\Delta t}\,e^{-\frac12\sigma_m^2\Delta t^2 T^2}.
\end{split}
\label{eq:Vslicewise-free}
\end{equation}
In momentum space, a successful slice multiplies amplitudes by
\begin{equation}
\psi(p)\ \longmapsto\
\exp\!\Big(-i\frac{p^2}{2m}\Delta t\Big)\;
\exp\!\Big(-\frac{\sigma_m^2\Delta t^2}{8m^2}\,p^4\Big)\,\psi(p).
\label{eq:Vslicewise-free-p}
\end{equation}
Thus, coherent mass-smearing induces a smooth \emph{$p^4$} UV suppressor.

In addition, one has
$V^{(T)}_{\sigma_m,\Delta t}=e^{-iT\Delta t}e^{-(\sigma_m^2\Delta t^2/2)T^2}$ and therefore
\begin{equation}
\big(V^{(T)}_{\sigma_m,\Delta t}\big)^\dagger V^{(T)}_{\sigma_m,\Delta t}
= e^{-\sigma_m^2\Delta t^2\,T^2}
= \exp\!\Big(-\frac{\sigma_m^2\Delta t^2}{4m^2}\,p^4\Big).
\end{equation}
Thus for an input state $\rho$ the per-slice success probability is
$p_1=\Tr\!\big[e^{-\sigma_m^2\Delta t^2T^2}\rho\big]$.
If success is required at each of $N=t/\Delta t$ slices, then in the free case
\begin{equation}
\begin{split}
p_N
&=\Tr\!\left[\left(\big(V^{(T)}_{\sigma_m,\Delta t}\big)^\dagger
V^{(T)}_{\sigma_m,\Delta t}\right)^N \rho\right] \\
&=\Tr\!\big[e^{-\sigma_m^2 t\,\Delta t\,T^2}\rho\big].
\end{split}
\label{eq:pn-mass}
\end{equation}
Under the diffusive scaling $\sigma_m^2=\kappa/\Delta t$, Eq.~\eqref{eq:pn-mass}
reduces to $p_N=\Tr\!\big[e^{-\kappa t\,T^2}\rho\big]$.

The short-time kernel is the convergent Fourier integral
\begin{multline}
K^{(T)}_{\sigma_m}(x',x;\Delta t)
=\int_{\mathbb R}\frac{dp}{2\pi}\;
e^{ip(x'-x)}\,e^{-i\frac{p^2}{2m}\Delta t}\\
\times\exp\!\Big(-\frac{\sigma_m^2\Delta t^2}{8m^2}\,p^4\Big).
\label{eq:kernel-p4}
\end{multline}
A convenient ``soft cutoff'' scale is obtained by setting the damping exponent to $1$:
\[
\frac{\sigma_m^2\Delta t^2}{8m^2}\,p_\ast^4\sim 1
\quad\Rightarrow\quad
p_\ast \sim \frac{m^{1/2}}{(\sigma_m\,\Delta t)^{1/2}}.
\]
\noindent
This estimate is accurate up to a factor $8^{1/4}$.
Hence the construction suppresses momenta $|p|\gtrsim p_\ast$ in each short-time slice.

\paragraph{Including a potential.}
For a general $V(\hat x)$, it is natural (and sufficient for the present discussion) to combine
\eqref{eq:Vslicewise-def} with a symmetric split-step approximation at small $\Delta t$,
\[
e^{-i(uT+V)\Delta t}
=
e^{-iV\Delta t/2}\,e^{-iuT\Delta t}\,e^{-iV\Delta t/2}
\;+\;O(\Delta t^3),
\]
so the $u$-integral acts only on the middle kinetic factor. This yields the practical short-time form
\begin{equation}
V^{(T)}_{\sigma_m,\Delta t}
\approx
e^{-iV\Delta t/2}\;
\Big(e^{-iT\Delta t}\,e^{-\frac12\sigma_m^2\Delta t^2 T^2}\Big)\;
e^{-iV\Delta t/2},
\label{eq:Vslicewise-Strang-fixed}
\end{equation}
which (i) reproduces the exact free-particle result when $V=0$ and (ii) makes transparent that the induced
suppression is a function of $T^2$ and therefore targets large momentum.

In the same spirit, a convenient kernel-level expression is
\begin{multline}
K^{(T)}_{\sigma_m}(x',x;\Delta t)
\approx
\exp\!\Big(-i\frac{\Delta t}{2}\big[V(x')+V(x)\big]\Big)
\int_{\mathbb R}\frac{dp}{2\pi}\; \\
e^{ip(x'-x)}\,e^{-i\frac{p^2}{2m}\Delta t}
\exp\!\Big(-\frac{\sigma_m^2\Delta t^2}{8m^2}\,p^4\Big),
\label{eq:kernel-p4-V}
\end{multline}
and for smooth $V$ one may further replace $\tfrac12(V(x')+V(x))\approx V(x_m)$ with $x_m=(x'+x)/2$.

\subsection{Accumulation over many slices}
\label{subsec:mass-scaling}

In the free case, since $T$ commutes with $T^2$, the $N=t/\Delta t$ step product can be evaluated exactly:
\begin{equation}
\begin{split}
\big(V^{(T)}_{\sigma_m,\Delta t}\big)^N
&=\Big(e^{-iT\Delta t}\,e^{-\frac12\sigma_m^2\Delta t^2T^2}\Big)^N \\
&=e^{-iTt}\,e^{-\frac12\sigma_m^2 t\,\Delta t\,T^2}.
\end{split}
\label{eq:Vslicewise-many}
\end{equation}
If the width of the per-slice fluctuation distribution is held fixed (fixed $\sigma_m$) and one refines the
slicing $\Delta t\to 0$ at fixed total time $t$, then the cumulative non-unitary factor
$e^{-(\sigma_m^2/2)\,t\Delta t\,T^2}$ disappears. In that regime the construction is best viewed as a
\emph{short-time regulator} of intermediate kernels, not as a finite modification of the continuous-time
unitary evolution.

If one nevertheless wishes to keep a finite damping in a continuous-time limit, one can adopt the diffusive scaling
\begin{equation}
\sigma_m^2=\frac{\kappa}{\Delta t},
\qquad \kappa>0\ \text{fixed},
\label{eq:diffusive-sigma}
\end{equation}
which turns \eqref{eq:Vslicewise-many} into the persistent non-Hermitian correction
\begin{equation}
\begin{split}
\big(V^{(T)}_{\sigma_m,\Delta t}\big)^N
&\to \exp\!\Big(-iTt-\frac{\kappa t}{2}\,T^2\Big) \\
&= \exp\!\Big(-i\frac{p^2}{2m}t-\frac{\kappa t}{8m^2}\,p^4\Big).
\end{split}
\label{eq:diffusive-p4}
\end{equation}
We emphasize that in the present work the conservative interpretation is the short-time one:
fixed $\sigma_m$ at finite slicing, with the regulator removable by $\sigma_m\to 0$.

\subsection{Conditioned ``branch selection'' and a classical-looking drift}
\label{subsec:mass-classical}

The $p^4$ factor derived above is most naturally interpreted at the level of the
normalized conditioned state (i.e.\ after a successful postselection), because the
normalization step makes the update nonlinear and can progressively \emph{select} a narrow
momentum sector.  In this subsection we make that selection mechanism explicit and connect it to
a classical-looking evolution of expectation values.

\paragraph{A classical evolution equation for the conditioned momentum distribution (free case).}
Take $V=0$.  After one successful slice, Eq.~\eqref{eq:Vslicewise-free-p} implies that the momentum
amplitudes are multiplied by $\exp[-(\sigma_m^2\Delta t^2/(8m^2))p^4]$, hence the momentum
probability density is reweighted as
\begin{equation}
\begin{split}
P_{n+1}(p)
&=\frac{P_n(p)\,W_{\Delta t}(p)}{\int dp\,P_n(p)\,W_{\Delta t}(p)},\\
W_{\Delta t}(p)
&:=\exp\!\Big(-\frac{\sigma_m^2\Delta t^2}{4m^2}\,p^4\Big).
\end{split}
\label{eq:bayes-update}
\end{equation}
where $n$ labels successful slices.  This is a simple ``Bayesian''/selection update: momenta with larger
$|p|$ are exponentially suppressed relative to smaller $|p|$.

Iterating \eqref{eq:bayes-update} for $N=t/\Delta t$ successful slices gives the closed form already implicit
in Eq.~\eqref{eq:Vslicewise-many}:
\begin{equation}
\begin{split}
P_t(p)
&=\frac{P_0(p)\,\exp\!\big[-\Gamma(t)\,p^4\big]}{\int dp\,P_0(p)\,\exp\!\big[-\Gamma(t)\,p^4\big]},\\
\Gamma(t)
&:=\frac{\sigma_m^2\,t\,\Delta t}{4m^2}.
\end{split}
\label{eq:Pt-closed}
\end{equation}
(Under the diffusive scaling $\sigma_m^2=\kappa/\Delta t$, one simply has $\Gamma(t)=\kappa t/(4m^2)$.)

Treating $t=n\Delta t$ as a coarse-grained continuous variable, Eq.~\eqref{eq:Pt-closed} implies the effective deterministic evolution equation
\begin{equation}
\begin{split}
\partial_t P_t(p)
&= -\dot\Gamma(t)\,\Big(p^4-\expval{p^4}_t\Big)\,P_t(p),\\
\dot\Gamma(t)
&=\frac{\sigma_m^2\,\Delta t}{4m^2}.
\end{split}
\label{eq:replicator}
\end{equation}
In the diffusive scaling one instead has $\dot\Gamma=\kappa/(4m^2)$.
This is a \emph{classical} (c-number) evolution equation: it governs a probability density by a
replicator/selection dynamics in momentum space.

\paragraph{Moment dynamics and monotonic decay of high-momentum content.}
From \eqref{eq:replicator}, any moment obeys
\begin{equation}
\begin{split}
\frac{d}{dt}\expval{p^k}_t
&= -\dot\Gamma(t)\,\Big(\expval{p^{k+4}}_t-\expval{p^k}_t\,\expval{p^4}_t\Big) \\
&= -\dot\Gamma(t)\,{\rm Cov}_t\!\big(p^k,p^4\big).
\end{split}
\label{eq:moment-eq}
\end{equation}
Two special cases make the ``classicalization'' toward low momentum transparent:
\begin{align}
\frac{d}{dt}\expval{p^4}_t
&= -\dot\Gamma(t)\,\Big(\expval{p^8}_t-\expval{p^4}_t^{\,2}\Big) \notag\\
&= -\dot\Gamma(t)\,{\rm Var}_t(p^4)\ \le\ 0,
\label{eq:p4-monotone}\\[4pt]
\frac{d}{dt}\expval{T^2}_t
&= -\frac{\dot\Gamma(t)}{4m^2}\,{\rm Var}_t(p^4)\ \le\ 0.
\label{eq:T2-monotone}
\end{align}
\noindent
Here $T=\frac{p^2}{2m}$.
Thus, in conditioned histories, the $p^4$-weight drives a monotonic reduction of the high-momentum content
(as measured by $\expval{p^4}$ or $\expval{T^2}$). The selection becomes strong once $\Gamma(t)\gg 1$, in which case
a typical surviving momentum scale is
\begin{equation}
p_\ast(t)\sim \Gamma(t)^{-1/4}.
\label{eq:pstar}
\end{equation}

\paragraph{Expectation value equations and an emergent classical branch.}
To connect to a ``classical trajectory,'' consider a spatially uniform vector potential $A(t)$
(in a gauge with vanishing scalar potential), so that the Hamiltonian is
\begin{equation}
H(t)=\frac{(p-A(t))^2}{2m},
\qquad
\dot x = \frac{p-A(t)}{m}.
\label{eq:min-coupling}
\end{equation}
For \emph{unitary} evolution under \eqref{eq:min-coupling}, the (canonical) momentum is conserved and
Ehrenfest’s theorem gives
\begin{equation}
\frac{d}{dt}\expval{x}_t=\frac{1}{m}\big(\expval{p}_t-A(t)\big),
\qquad
\frac{d}{dt}\expval{p}_t=0.
\label{eq:ehrenfest-unitary}
\end{equation}
In the present protocol, the Hamiltonian part still generates the usual drift
$d\expval{x}_t/dt=(\expval{p}_t-A(t))/m$, but the \emph{conditioning} changes the momentum statistics according
to \eqref{eq:replicator}. In particular,
\begin{equation}
\begin{split}
\frac{d}{dt}\expval{p}_t
&= -\dot\Gamma(t)\,\Big(\expval{p^{5}}_t-\expval{p}_t\,\expval{p^4}_t\Big) \\
&= -\dot\Gamma(t)\,{\rm Cov}_t(p,p^4).
\end{split}
\label{eq:p-drift}
\end{equation}
Whenever the state has a nonzero momentum spread, ${\rm Cov}_t(p,p^4)$ is generically nonzero and the
conditioning produces a genuine drift of the mean momentum.  For a wavepacket whose momentum distribution is
approximately Gaussian with mean $\bar p(t)$ and variance $(\Delta p_t)^2$, a short moment expansion gives
\begin{equation}
\frac{d\bar p}{dt}\approx -\dot\Gamma(t)\,
\Big(4\,\bar p^{\,3}(\Delta p_t)^2 + 12\,\bar p\,(\Delta p_t)^4\Big).
\label{eq:pbar-gauss}
\end{equation}
so (for $\bar p>0$) the mean momentum is driven downward toward $0$; similarly for $\bar p<0$ it is driven upward.
Combining \eqref{eq:ehrenfest-unitary} with this drift yields a closed, classical-looking description of the
wavepacket center:
\begin{equation}
\begin{split}
\frac{d}{dt}\expval{x}_t&=\frac{1}{m}\big(\bar p(t)-A(t)\big),\\
\frac{d\bar p}{dt}&=-\dot\Gamma(t)\,{\rm Cov}_t(p,p^4)
\quad (\text{or }\approx\text{ \eqref{eq:pbar-gauss}}).
\end{split}
\label{eq:classical-looking-system}
\end{equation}
As the conditioned distribution narrows and $\bar p(t)\to 0$, Eq.~\eqref{eq:classical-looking-system} approaches the
\emph{classical $p=0$ branch} of \eqref{eq:min-coupling}:
\begin{equation}
\expval{x}_t\ \simeq\ \expval{x}_0 - \frac{1}{m}\int_0^t A(s)\,ds.
\label{eq:p0-branch}
\end{equation}

\paragraph{Which ``branch'' is selected depends on how the fluctuation enters.}
The discussion above corresponds to the choice in Eq.~\eqref{eq:Hu-def}, where the fluctuating prefactor multiplies
the canonical kinetic term $p^2/(2m)$ and therefore the induced filter is centered at $p=0$.
If instead one chooses a gauge-covariant fluctuation $H(u;t)=u\,(p-A(t))^2/(2m)$, then the same mechanism produces a
filter centered at small \emph{kinetic} momentum $(p-A(t))\simeq 0$, i.e.\ it favors a different semiclassical branch.
For constant $A$ this corresponds to selecting $\dot x\simeq 0$; for rapidly varying $A(t)$ the success probability can
become small because exact tracking $(p-A(t))\approx 0$ conflicts with the conservation of the canonical $p$.

\begin{remark}[Unconditioned channel (for comparison)]
If the control is discarded (no postselection), the system undergoes a trace-preserving channel rather than a
Kraus filter. In the free case,
\[
\Phi^{(T)}_{\sigma_m,\Delta t}(\rho)=\int du\,g_{\sigma_m}(u)\;e^{-iuT\Delta t}\,\rho\,e^{+iuT\Delta t},
\]
which dephases in the kinetic-energy basis but does not reweight the momentum distribution. Under the diffusive
scaling $\sigma_m^2=\kappa/\Delta t$ one obtains the standard double-commutator form
$\dot\rho=-i[T,\rho]-(\kappa/2)[T,[T,\rho]]$.
\end{remark}

\paragraph{Takeaway.}
In the conditioned (postselected) histories, coherent mass-smearing induces a momentum-selection dynamics
\eqref{eq:replicator} that monotonically reduces high-momentum content \eqref{eq:p4-monotone} and can drive
expectation values toward a classical-looking branch such as $p\simeq 0$ in \eqref{eq:p0-branch}.
At fixed $\sigma_m$ the strength $\dot\Gamma(t)\propto\Delta t$ vanishes as $\Delta t\to 0$, consistent with
Sec.~\ref{subsec:mass-scaling}; a persistent continuous-time selection corresponds to the diffusive scaling
$\sigma_m^2=\kappa/\Delta t$.

\subsection{Feynman sum over paths}
The ``branch selection'' described above can also be seen directly at the level of the
\emph{conditioned propagator} (i.e.\ the transition amplitude given success of the postselection at each slice).
For concreteness, consider the free case $V=0$ and define the conditioned $N$-slice propagator
\begin{equation}
\begin{split}
K^{\rm(cond)}_{\sigma_m}(x_f,x_i;t)
&:=\mel{x_f}{\big(V^{(T)}_{\sigma_m,\Delta t}\big)^N}{x_i},\\
N&=\frac{t}{\Delta t}.
\end{split}
\label{eq:Kcond-def}
\end{equation}
Using the short-time kernel \eqref{eq:kernel-p4} and inserting intermediate positions
$\mathbbm{1}=\int dx_n\,\ket{x_n}\!\bra{x_n}$ at each slice, one obtains a time-sliced
Feynman ``sum over paths'' representation,
\begin{multline}
K^{\rm(cond)}_{\sigma_m}(x_f,x_i;t)=\\
\int\!\!\Big(\prod_{n=1}^{N-1}dx_n\Big)
\prod_{n=0}^{N-1}\Bigg[
\int_{\mathbb R}\frac{dp_n}{2\pi}\;
 e^{i p_n(x_{n+1}-x_n)}\\
\times e^{-i\Delta t\,p_n^2/(2m)}
\exp\!\Big(-\frac{\sigma_m^2\Delta t^2}{8m^2}\,p_n^4\Big)
\Bigg].
\label{eq:Kcond-phase-space}
\end{multline}
with $x_0=x_i$ and $x_N=x_f$.
Equation \eqref{eq:Kcond-phase-space} is the usual phase-space path integral weight
$e^{i\sum[p\Delta x-\Delta t\,H(p)]}$, multiplied by an additional \emph{positive} factor that suppresses
large momenta in each slice.

Under the diffusive scaling $\sigma_m^2=\kappa/\Delta t$, the product of suppressors has a clean continuum form:
\begin{equation}
\begin{split}
\prod_{n=0}^{N-1}\exp\!\Big(-\frac{\sigma_m^2\Delta t^2}{8m^2}\,p_n^4\Big)
&= \exp\!\Big(-\frac{\kappa}{8m^2}\sum_{n=0}^{N-1}\Delta t\,p_n^4\Big) \\
&\longrightarrow \exp\!\Big(-\frac{\kappa}{8m^2}\int_0^t ds\,p(s)^4\Big).
\end{split}
\label{eq:p4-functional}
\end{equation}
Thus, in this scaling the conditioned propagator can be viewed as a Feynman sum over phase-space paths with a
\emph{complex effective action}
\begin{equation}
\begin{split}
K^{\rm(cond)}_{\kappa}(x_f,x_i;t)
\sim \int \mathcal Dp\,\mathcal Dx\;
\exp\!\Bigg\{
&i\int_0^t ds\Big[p\,\dot x-\frac{p^2}{2m}\Big] \\
&-\frac{\kappa}{8m^2}\int_0^t ds\,p^4
\Bigg\}.
\end{split}
\label{eq:Seff}
\end{equation}
The extra term is precisely the interference-induced filter: it provides an \emph{exponential} penalty for paths
with large $|p(s)|$, making the path integral ``Laplace-like'' in momentum space.

\paragraph{Dominant path in the strong-selection limit.}
When the accumulated damping is strong (heuristically, when $\kappa t/m^2$ is large, equivalently when the
momentum scale \eqref{eq:pstar} becomes small), the integral \eqref{eq:Seff} is dominated by trajectories that
minimize the positive functional $\int_0^t ds\,p(s)^4$ subject to the boundary constraint on the net displacement.
For the free particle, $x_f-x_i=\int_0^t ds\,\dot x(s)$ and the phase-space term is stationary at $\dot x=p/m$
in the usual semiclassical sense. Combining these, the relevant constraint becomes
\begin{equation}
\int_0^t ds\,p(s) \;=\; m(x_f-x_i).
\label{eq:constraint}
\end{equation}
Among all functions $p(s)$ satisfying \eqref{eq:constraint}, the quantity $\int_0^t ds\,p(s)^4$ is minimized by a
\emph{constant} momentum $p(s)\equiv p_{\rm cl}$ (this is a direct consequence of convexity/Jensen's inequality).
Therefore the least-suppressed contribution comes from
\begin{equation}
p_{\rm cl}=\frac{m(x_f-x_i)}{t},
\qquad
x_{\rm cl}(s)=x_i+\frac{x_f-x_i}{t}\,s,
\label{eq:free-classical-path}
\end{equation}
namely the standard classical straight-line path of the free particle.
Fluctuations away from this trajectory increase $\int_0^t ds\,p(s)^4$ and are exponentially suppressed by
\eqref{eq:p4-functional}, so in the strong-selection regime the conditioned propagator is dominated by a narrow
tube of paths around \eqref{eq:free-classical-path}.

\paragraph{Connection to the ``$p\simeq 0$ branch'' viewpoint.}
The discussion above fixes both endpoints $x_i,x_f$ and therefore selects the classical constant-momentum path
compatible with that displacement.
If instead one considers an initial wavepacket and asks which histories are favored by conditioning (as in the
replicator dynamics \eqref{eq:replicator}), the same weight \eqref{eq:p4-functional} suppresses all histories with
large momentum, driving the conditioned ensemble toward small $|p|$ and hence toward the $p\simeq 0$ branch when
that branch is compatible with the boundary data.
In the uniform vector-potential example \eqref{eq:min-coupling}, selecting $p\simeq 0$ yields the corresponding
classical branch \eqref{eq:p0-branch}. In this sense, the ``classical-looking drift'' in expectation values and the
dominance of a classical path in the Feynman sum are two complementary views of the same mechanism: the
interference-induced filter turns the conditioned sum over histories into one in which a small set of
(classically interpretable) trajectories dominate.

\section{Applications in singular quantum mechanics}

We now return to the main Gaussian \emph{energy-filter} construction of Sec.~2 and analyze the regulated short-time kernel
\begin{equation}
\begin{split}
K_\sigma(x',x;\Delta t)
&:=\matrixel{x'}{e^{-iH\Delta t}\,e^{-\frac{1}{2}\sigma^2\Delta t^2 H^2}}{x},\\
&\qquad \Delta t>0\ \text{(short time)}.
\end{split}
\label{eq:Ksigma-def}
\end{equation}
and apply it to three singular settings.  The extra factor
$e^{-\frac{1}{2}\sigma^2\Delta t^2 H^2}$ is a \emph{Gaussian energy filter}:
for any spectral value $E$ it contributes a multiplicative weight
$\exp\!\big(-\tfrac12\sigma^2\Delta t^2E^2\big)$, so \emph{both} large positive and large
negative energies are exponentially suppressed at each short step.

Throughout this section the underlying Hamiltonians are taken to be self-adjoint (or their standard
self-adjoint extensions), so the exact propagator $e^{-iHt}$ is well-defined. What becomes ill-behaved
in practice is the \emph{short-time approximation} used in time-sliced/path-integral representations
(e.g.\ midpoint approximations) for singular potentials. The filter in \eqref{eq:Ksigma-def} supplies an
explicit damping that renders these short-time kernels well behaved at fixed $\sigma>0$, and the
original unitary dynamics is recovered by sending $\sigma\to 0$ after the relevant short-time
manipulations.

\subsection{The kernel picture}

Let $\ket{E}$ denote a spectral basis of $H$ (including both discrete and continuous parts) with
$H\ket{E}=E\ket{E}$ and resolution of the identity
\(
\int d\nu(E)\,\ket{E}\!\bra{E}=\mathbbm{1},
\)
where $d\nu(E)$ stands for ``sum over bound states + integral over continuum''.
Writing $\varphi_E(x):=\braket{x}{E}$, a single regulated step acts diagonally in energy:
\[
\braket{E}{\psi}\ \longmapsto\ e^{-iE\Delta t}\,e^{-\tfrac12\sigma^2\Delta t^2E^2}\,\braket{E}{\psi}.
\]
Correspondingly, the regulated short-time kernel is the familiar spectral superposition
\begin{equation}
\begin{split}
K_\sigma(x',x;\Delta t)
&=\int d\nu(E)\; e^{-iE\Delta t}e^{-\tfrac12\sigma^2\Delta t^2E^2} \\
&\qquad\times \varphi_E(x')\,\varphi_E^\ast(x).
\end{split}
\label{eq:Ksigma-spectral}
\end{equation}
The key point is the Gaussian factor: it suppresses $|E|\gtrsim 1/(\sigma\Delta t)$ so strongly that,
for the Schr\"odinger operators considered below (where the spectral density and eigenfunctions have
at most polynomial growth in $|E|$), the high-energy tail is effectively cut off and the short-time
kernel is well behaved for any fixed $\sigma>0$ and $\Delta t>0$.

To connect with the standard time-sliced path integral, write $H=T+V$ with $T=p^2/2m$.
Using the midpoint/symmetric split-step approximation for the unitary part and keeping the leading
potential-dominated piece of $H^2$ at the midpoint $x_m=\tfrac12(x'+x)$ gives, for small $\Delta t$,
\begin{equation}
\begin{split}
K_\sigma(x',x;\Delta t)
&\approx K_0(x',x;\Delta t)\,\exp\!\Big(-i\,V(x_m)\Delta t\Big) \\
&\qquad\times \exp\!\Big(-\tfrac{1}{2}\sigma^2\Delta t^2\,V(x_m)^2\Big).
\end{split}
\label{eq:parametrix}
\end{equation}
with $K_0(x',x;\Delta t)=(\frac{m}{2\pi i\Delta t})^{1/2}
\exp\!\big(\frac{im(x'-x)^2}{2\Delta t}\big)$ (one spatial dimension).
The last exponential in \eqref{eq:parametrix} is the \emph{local damping}: in a single slice,
regions where $|V|$ is large (including singular regions) are suppressed by
$\exp\!\big(-\tfrac12\sigma^2\Delta t^2V(x_m)^2\big)$.
Corrections from $T^2$ and $TV+VT$ generate derivative terms; for the singular examples below,
the displayed $V^2$ contribution captures the dominant short-distance suppression.

Finally, because both factors $e^{-iH\Delta t}$ and $e^{-\frac12\sigma^2\Delta t^2H^2}$ are functions
of the same operator $H$, they commute. Hence after $N=t/\Delta t$ steps one has the exact identity
\[
\big(e^{-iH\Delta t}\,e^{-\tfrac12\sigma^2\Delta t^2H^2}\big)^N
= e^{-iHt}\,e^{-\tfrac12\sigma^2 t\,\Delta t\,H^2}.
\]
In particular, at fixed $\sigma$ the regulator disappears as $\Delta t\to 0$
(Prop.~\ref{prop:continuum}), while under diffusive scaling $\sigma^2=\kappa/\Delta t$ it yields the
persistent damping $e^{-(\kappa t/2)H^2}$.

\subsection{Outside-negative ``infinite well'' (singular \texorpdfstring{$-\infty$}{-∞} outside)}

Take the step potential
\[
V_{V_0}(x)=
\begin{cases}
0,& x\in[0,a],\\[2pt]
-\,V_0,& x\notin[0,a],
\end{cases}
\qquad V_0>0,\quad V_0\to\infty,
\]
and denote the corresponding Hamiltonian by $H_{V_0}=T+V_{V_0}$.
For each fixed $V_0$ the operator is bounded below by $-V_0$, but in the limit $V_0\to\infty$
the bottom of the spectrum drifts to $-\infty$, so naive short-time propagation develops an
uncontrolled large-negative-energy sector.

It is helpful to distinguish \emph{geometric} (position) projections from \emph{spectral} ones.
Let
\[
\Pi_{\mathrm{out}}:=\chi_{\mathbb R\setminus[0,a]}(\hat x),
\qquad
\Pi_{\mathrm{in}}:=\chi_{[0,a]}(\hat x)=\mathbbm{1}-\Pi_{\mathrm{out}},
\]
so that $(\Pi_{\mathrm{out}}\psi)(x)=\psi(x)$ for $x\notin[0,a]$ and $0$ otherwise.
For an energy cutoff $E_->0$ we write the spectral projector as
\[
P_{\le -E_-}(H_{V_0}) := \chi_{(-\infty,-E_-]}(H_{V_0}).
\]
A natural way to isolate energies that become arbitrarily negative as $V_0\to\infty$ is to choose
$E_-=\alpha V_0$ with any fixed $0<\alpha<1$.

Then the Gaussian filter suppresses that sector in operator norm:
for any state $\psi$,
\begin{multline}
\big\|e^{-iH_{V_0}\Delta t}e^{-\tfrac12\sigma^2\Delta t^2 H_{V_0}^2}
P_{\le -\alpha V_0}(H_{V_0})\psi\big\| \\
\le e^{-\tfrac12\sigma^2\Delta t^2 (\alpha V_0)^2}\,\|\psi\|.
\label{eq:neg-energy-suppression}
\end{multline}
This is immediate from the spectral theorem: on $\mathrm{Ran}\,P_{\le -\alpha V_0}$ one has
$H_{V_0}^2\ge (\alpha V_0)^2$, while $e^{-iH_{V_0}\Delta t}$ is unitary.

The position-space short-time approximation \eqref{eq:parametrix} shows the same mechanism even more directly.
Whenever the midpoint $x_m$ lies outside $[0,a]$, we have $V(x_m)=-V_0$ and the local damping
factor becomes $e^{-\frac12\sigma^2\Delta t^2V_0^2}$, so every excursion into the outside region is
exponentially suppressed at the level of a \emph{single} slice.
Thus, at fixed $\sigma>0$ and $\Delta t>0$, taking $V_0\to\infty$ selects a well-defined
short-time kernel supported on paths that remain in $[0,a]$ throughout the slicing, avoiding any
``runaway'' contribution from the emerging $-\infty$ spectral tail.
As $\sigma\to 0$ (after the appropriate continuum/renormalization limit), ordinary unitary
dynamics is recovered.

\subsection{One-dimensional Coulomb singularity}

For
\[
H_g=-\frac{1}{2m}\partial_x^2+\frac{g}{|x|},
\]
the coordinate singularity at $x=0$ makes the naive midpoint short-time kernel ill-behaved,
especially for even states. Nevertheless, $H_g$ is a standard self-adjoint Hamiltonian (as a form
sum; see \cite{Kato1995,ReedSimonII1975}).

The regulated step multiplies each energy component by
$e^{-\frac12\sigma^2\Delta t^2E^2}$, uniformly controlling the high-energy part that drives the
short-time singular behavior.  In the same midpoint approximation as \eqref{eq:parametrix}, one
finds
\[
\begin{aligned}
K_\sigma(x',x;\Delta t)
&\approx K_0(x',x;\Delta t)\,\exp\!\Big(-i\,\tfrac{g\Delta t}{|x_m|}\Big)\\
&\qquad\times \exp\!\Big(-\tfrac{\sigma^2\Delta t^2 g^2}{2\,x_m^2}\Big).
\end{aligned}
\]
The crucial point is that the regulator produces an \emph{integrable} suppression near the origin:
for any fixed $\delta>0$,
\[
\sup_{|x_m|<\delta}|K_\sigma(x',x;\Delta t)|
\lesssim \Big(\tfrac{m}{2\pi\Delta t}\Big)^{1/2}
\exp\!\Big(-\tfrac{\sigma^2\Delta t^2 g^2}{2\,\delta^2}\Big).
\]
and, more strongly, the weight $e^{-c/x_m^2}$ makes the near-origin contribution to any
intermediate-$x$ integration finite (and in fact super-exponentially small).
Indeed, for any $c>0$ one has $\int_0^\delta dx\,e^{-c/x^2}<\infty$, so the regulated midpoint weight
renders the near-origin contribution to the time-sliced integral manifestly finite.
Equivalently, the ``danger zone'' in a single slice is parametrically
\(
|x_m|\lesssim \sigma\,\Delta t\,|g|,
\)
and is therefore squeezed away in the $\Delta t\to 0$ continuum limit at fixed $\sigma$
(Prop.~\ref{prop:continuum}).  At the same time, for any fixed $\sigma>0$ the regulated kernel
remains finite at short times.

\subsection{Spiked oscillators \texorpdfstring{$V(x)=\eta|x|^{-\nu}$}{V = η|x|^{−ν}}}

For $0<\nu\le 2$, the spike $|x|^{-\nu}$ diverges at $x=0$ but the corresponding Schr\"odinger operator
is nevertheless well-defined and bounded below (Hardy’s inequality for $\nu=2$ and simple comparison
bounds for $\nu<2$); see, e.g., \cite{Kato1995,ReedSimonII1975,Hardy1920}.

As in \eqref{eq:Ksigma-spectral}, the Gaussian factor yields an energy-space filter, while in
position space \eqref{eq:parametrix} gives the explicit local damping
\[
\begin{aligned}
K_\sigma(x',x;\Delta t)
&\approx K_0(x',x;\Delta t)\,\exp\!\big(-i\,\eta |x_m|^{-\nu}\Delta t\big)\\
&\qquad\times \exp\!\Big(-\tfrac12\sigma^2\Delta t^2\eta^2\,|x_m|^{-2\nu}\Big).
\end{aligned}
\]
Thus, in a single slice, paths that approach the spike closer than the scale
\[
|x_m|\ \lesssim\ \Big(\tfrac{\sigma\,\Delta t\,|\eta|}{\sqrt{2}}\Big)^{\!1/\nu}
\]
are exponentially suppressed.  In particular, if one wishes to render the short-time kernel
insensitive to the spike below a target spatial resolution $r_{\rm sing}$, a convenient rule of
thumb is
\[
\sigma\,\Delta t \ \gtrsim\  \frac{\sqrt{2}\,r_{\rm sing}^{\nu}}{|\eta|}.
\]
As in the Coulomb case, the suppressed region shrinks with $\Delta t$ at fixed $\sigma$, so the
continuum limit recovers the standard unitary dynamics while the regulated short-time kernel
remains well-defined for any fixed $\sigma>0$.

For orientation, it is useful to recall two standard spectral facts for the spiked oscillator
\(
H_\nu=-\frac{1}{2m}\partial_x^2+\frac12 m\omega^2x^2+\eta|x|^{-\nu}.
\)
Basic energy estimates imply that adding the harmonic term makes the spectrum discrete, so the regulated step is even
Hilbert--Schmidt (its squared HS norm is $\sum_n e^{-\sigma^2\Delta t^2 E_n^2}<\infty$). This justifies using the regulated kernel without functional-analytic detours; see \cite{Kato1995,ReedSimonII1975,Hardy1920} for background if desired.

\section{A local interference-based regulator for scalar QFT}

We now develop a field-theoretic analog of the interference-based mechanism introduced above. In this setting, the construction is used at the level of the functional integral: a local Gaussian averaging of the quartic coupling generates an additional local term in the effective action. Although the averaging can be written starting from the Minkowski amplitude, its role as a regulator is clearest after Wick rotation, where the induced term becomes a positive local $\phi^8$ interaction that suppresses large-field configurations and makes the stability of the Euclidean measure manifest.

\subsection{Local coupling average and corrected effective action}
Start from the Minkowski Lagrangian with quartic interaction
\begin{equation}
\mathcal{L}_\lambda = \frac{1}{2}(\partial \phi)^2 - \frac{1}{2} m^2 \phi^2 - \lambda\, \phi^4.
\end{equation}
We promote the coupling to a \emph{local} variable $\lambda(x)$ and implement a \emph{local coherent
smearing} by integrating over $\lambda(x)$ in the \emph{amplitude} with a Gaussian functional weight,
\begin{multline}
\int \mathcal{D}\lambda\;
\exp\!\left[-\int d^d x\,\frac{(\lambda(x)-\lambda_0)^2}{2\sigma^2}\right]\\
\times \exp\!\left[-i\int d^d x\,\lambda(x)\phi^4(x)\right].
\label{eq:lambda-functional}
\end{multline}
(Equivalently, one may write $\lambda(x)=\lambda_0+\sigma\,\xi(x)$ with $\xi$ a Gaussian white-noise
field satisfying $\langle\xi(x)\xi(y)\rangle=\delta^{(d)}(x-y)$. A strictly local prior is thus most
naturally understood with an implicit UV regulator, e.g.\ a lattice.)

Because the weight in \eqref{eq:lambda-functional} is ultralocal, the $\lambda$-integral factorizes
pointwise and yields the Gaussian characteristic function, producing the effective replacement
\begin{multline}
\exp\!\left[-i\int d^d x\,\lambda(x)\phi^4(x)\right] \\
\longmapsto \exp\!\left[-i\int d^d x\,\lambda_0\,\phi^4(x)\right]
\exp\!\left[-\frac{\sigma^2}{2}\int d^d x\,\phi^8(x)\right].
\label{eq:gauss-characteristic}
\end{multline}
up to an overall $\phi$-independent normalization constant.

\medskip

\noindent\emph{Scope of the regulator.} The induced $+\frac{\sigma^2}{2}\phi^8$ term is a \emph{large-field}
stabilizer: it suppresses configurations with large $|\phi|$ and guarantees a positive even-polynomial
potential in Euclidean signature. For $\lambda_0>0$ the Euclidean $\phi^4$ theory is already bounded below,
so the extra term is not required for stability, but it provides an additional tail suppression that can be
convenient in approximate treatments. More importantly, it renders the Euclidean measure stable even when
$\lambda_0<0$ (or when negative quartic terms arise in effective actions). By contrast, the usual short-distance
UV divergences of correlation functions remain and must be treated with a conventional UV regulator (lattice,
cutoff, dimensional regularization) and renormalization at fixed $\sigma$, after which one takes $\sigma\to 0$.

Therefore, the effective Minkowski action can be written as
\begin{equation}
\begin{split}
S_{\text{eff}}[\phi]
=\int d^d x \Big[&\frac{1}{2}(\partial \phi)^2
- \frac{1}{2} m^2 \phi^2 - \lambda_0 \phi^4(x) \\
&+ i\,\frac{\sigma^2}{2}\,\phi^8(x)\Big].
\end{split}
\label{eq:phi8-minkowski}
\end{equation}
so that $e^{iS_{\text{eff}}}$ contains the damping factor
$\exp\{-\frac{\sigma^2}{2}\int d^d x\, \phi^8(x)\}$.

After Wick rotation to Euclidean signature, this corresponds to the local Euclidean action
\begin{equation}
\begin{split}
S_{E,\sigma}[\phi] = \int d^d x \Big[&\frac{1}{2}(\partial \phi)^2
+ \frac{1}{2} m^2 \phi^2 + \lambda_0 \phi^4(x) \\
&+ \frac{\sigma^2}{2}\,\phi^8(x)\Big].
\end{split}
\label{eq:phi8-euclidean}
\end{equation}
In particular, for any fixed $\sigma>0$ the large-field tail is stabilized by the positive $\phi^8$
term, while the limit $\sigma\to 0$ recovers the target $\phi^4$ theory.

\begin{remark}[Local vs.\ global smearing and OS positivity]
Averaging a \emph{global} coupling produces a nonlocal $(\int \phi^4)^2$ term and obscures reflection positivity. The \emph{local} smearing above yields a positive even-polynomial potential in Euclidean signature and sits naturally within the Osterwalder--Schrader framework \cite{OS1973,OS1975}.
\end{remark}

\subsection{Dimensions, stability, and renormalization at fixed $\sigma$}
In $d=4$ one has $[\phi]=1$ and $[\phi^8]=8$, so $[\sigma^2]=-4$. Writing $\sigma^2 = c\,\Lambda^{-4}$ makes the regulator an \emph{irrelevant} local operator suppressed by a UV scale $\Lambda$. At a renormalization/external scale $\mu$, the associated dimensionless coupling is
$g_8(\mu)\sim \sigma^2\mu^4\sim c(\mu/\Lambda)^4$, so contributions of the induced operator to IR observables
are parametrically suppressed in $d=4$.
In Euclidean signature the additional potential $+\frac{\sigma^2}{2}\phi^8$ is positive and stabilizes the large-field tails of the measure. Perturbation theory and renormalization are carried out at fixed $\sigma$, after which one sends $\sigma\to 0$ (or moves to the IR where the induced operator is negligible) to recover the target theory.

\section{Relation to existing regularization and measurement literature\label{sec:relation-existing-methods}}

Having developed the construction, we now place it in
relation to several neighboring frameworks. The comparisons
below clarify which aspects of the present regularizers are
standard: smooth spectral damping, nonunitary conditional
evolution, absorbing suppression, and large-field stabilization
all have long histories. They also isolate the specific point
of the present work: these structures arise here from an
interference-generated, postselected Kraus map, with the
regulator treated conservatively as removable.

\noindent\textit{Linear combinations of unitaries.}
At the algebraic level our conditioned map
\(K=\sum_j a_j U_j\) is a linear combination of unitary evolutions, and is therefore closely related to the linear-combination-of-unitaries (LCU) primitive used in quantum algorithms and Hamiltonian simulation \cite{ChildsWiebe2012,Berry2015,LowChuang2017}.  The similarity is important: both settings use an ancilla/control system, coherent branching into alternatives, and a postselection or projection step that realizes a non-unitary linear combination before normalization.  The difference is the goal.  In the LCU literature the linear combination is typically an algorithmic subroutine whose nonunitarity is an implementation overhead to be controlled, amplified, or embedded into a near-deterministic simulation of a target unitary.  In the present work the trace-decreasing part of the Kraus operator is the object of interest: for a narrow Gaussian superposition of nearby time translations it produces the explicit filter
\(e^{-(\sigma^2\Delta t^2/2)H^2}\).  Thus we use LCU-like coherent control not primarily to speed up Hamiltonian simulation or optimize query complexity, but to give an operational derivation of a removable short-time regulator.  Equivalently, the loss of norm is not interpreted as an algorithmic failure probability to be eliminated; it is the heralding probability associated with the regulator and is analyzed as part of the construction.

\noindent\textit{Continuous monitoring and quantum trajectories.}
The nonunitary conditional evolution also has a natural connection with continuous-time measurement theory and quantum trajectories \cite{BarchielliBelavkin1991,JacobsSteck2006,WisemanMilburn2009,BarchielliGregoratti2009}.  In particular, in the diffusive scaling discussed in Sec.~II, the product of Gaussian time-smearing steps has the same formal structure as an evolution with a non-Hermitian damping term proportional to \(H^2\), and this resembles the damping of components selected by continuous monitoring of an observable.  The difference is again physical and operational.  Continuous measurement theory describes repeated weak measurements of the system, stochastic measurement records, information gain, and measurement backaction; the corresponding unconditioned dynamics is typically a trace-preserving decohering channel.  By contrast, our regulator is obtained by coherent interference between alternative evolutions followed by a single heralding projection on the control.  No continuous measurement record of the system observable is used.  Moreover, the main regime in this paper is fixed \(\sigma\) at finite short-time slicing, followed by removal of the regulator, whereas the diffusive scaling is included only to display the possible finite non-Hermitian limit.

\noindent\textit{Spectral and heat-kernel regularization.}
The Gaussian factor \(e^{-(\sigma^2\Delta t^2/2)H^2}\) is also a smooth spectral cutoff.  This places it near the broad heat-kernel and spectral-regularization tradition, where factors such as \(e^{-sD}\), spectral cutoffs, and complex powers of elliptic operators are used to control short-distance behavior, one-loop divergences, anomalies, and spectral asymptotics \cite{Seeley1967,Vassilevich2003}.  Functionally, our energy filter may be viewed as a heat-kernel-type factor for the positive operator \(H^2\).  The differences are that (i) the filter is derived as the Fourier transform of a real-time distribution of time translations rather than inserted as an auxiliary Euclidean proper-time regulator, (ii) its scale is tied to the time-slice and the width of the coherent control distribution, and (iii) it damps large positive and large negative spectral values symmetrically, which is useful in the singular quantum-mechanical examples where large negative sectors can appear.  In the QFT part, the induced \(\phi^8\) term should likewise not be read as a replacement for standard ultraviolet regularization; it is a large-field stabilizer to be used together with an ordinary UV regulator and removed after renormalization.

\noindent\textit{Non-Hermitian damping and absorbing potentials.}
For finite damping, the regulated step can be rewritten as evolution generated by an effective non-Hermitian operator, and it is therefore natural to compare it with non-Hermitian quantum mechanics and with complex absorbing potentials \cite{Moiseyev2011,RissMeyer1995,Muga2004}.  The similarity is the deliberate use of nonunitary propagation to suppress unwanted components.  Standard absorbing potentials, however, are usually introduced as complex spatial potentials, often localized near artificial boundaries, in order to mimic outgoing-wave conditions, avoid reflections from a finite numerical grid, or compute resonance data.  The present damping is not an absorbing boundary condition and is not chosen phenomenologically to model loss into an external channel.  It is a positive spectral filter, or in the kinetic variant a \(p^4\) filter, arising from a unitary dilation followed by postselection.  Its trace decrease is accompanied by an explicit success probability, and the target unitary dynamics is recovered by removing the regulator.  Thus the construction is closer to a heralded regularizing Kraus map than to a non-Hermitian Hamiltonian taken as the fundamental generator.

\noindent\textit{Large-field stabilization and constructive functional integrals.}
The scalar-field part connects most directly to the constructive and Euclidean functional-integral literature, where stability of the Euclidean measure, ultraviolet cutoffs, reflection positivity, and renormalization are central issues \cite{Simon1974,GlimmJaffe1987,Rivasseau1991,OS1973,OS1975}.  Adding a positive higher even power of the field is a familiar way to control large-field tails, and from a Wilsonian viewpoint such higher-dimensional operators are irrelevant at sufficiently low scales \cite{WilsonKogut1974}.  Our local Gaussian average over the quartic coupling produces precisely such a positive \((\sigma^2/2)\phi^8\) contribution.  The difference from the constructive program is that we are not claiming a new nonperturbative construction of the continuum theory and we do not eliminate the need for a conventional UV cutoff and renormalization.  Rather, the local coupling average gives an operational route to a symmetry-compatible large-field stabilizer.  Renormalization is to be performed at fixed \(\sigma\), with the usual UV regulator in place, and the limit \(\sigma\to0\) is then taken to recover the target theory.

In summary, the proposed regulator overlaps with several established methods in its mathematical appearance, but differs in its origin and intended use.  It is an interference-generated, postselected Kraus map whose nonunitarity is a controlled and removable regularizing feature, not a claim that regularization, heat-kernel damping, absorbing potentials, or constructive stabilization are themselves new.

\section{Summary and outlook}
Our interference-based scheme yields three complementary, symmetry-respecting regularizers: an
\emph{energy-space} Gaussian filter from time-smearing in QM, a \emph{momentum-space} $p^4$ suppressor from coherent
mass fluctuations in QM, and a \emph{field-amplitude} suppressor from local coupling averaging in QFT. At the level
of functional form these are close cousins of familiar smooth regulators (spectral/heat-kernel filtering in QM and
higher-dimensional operators in Wilsonian effective actions), but here they arise from a concrete coherent-control
and postselection protocol and thus come with a direct operational interpretation of the regulator strength. In this
preliminary work we treat all three as regulators that are removed after renormalization. We also stress that our
regulator does not replace the program of self-adjoint extensions in singular QM; rather, it singles out a benign
short-time kernel that reproduces standard unitary dynamics as $\sigma\to 0$.

We have presented an interference-based regularization framework that (i) admits a clear short-time error bound showing recovery of unitary dynamics for narrow superpositions, (ii) yields in QM both a universal Gaussian energy filter from time-smearing and a closely related $p^4$ momentum suppressor from coherent mass fluctuations as a complementary short-time variant, and (iii) produces a corrected, local, symmetry-preserving scalar-field regulator with a positive $\phi^8$ term in Euclidean signature. The regulator can be used at fixed strength, standard renormalization performed, and then removed to restore the target theory. The conceptual contribution is the operational derivation: the regularizing factors appear as the
conditional map produced by a coherent superposition of evolutions, providing a physical motivation
for inserting familiar spectral/large-field filters in a locality- and symmetry-respecting way.

From a renormalization viewpoint, $\sigma$ can be treated as an additional smooth regulator knob rather than a
new fundamental constant. In QFT one keeps a conventional UV regulator, includes the induced local interaction
$(\sigma^2/2)\phi^8$ in the bare action, renormalizes at fixed $\sigma$ by matching a set of observables at a
scale $\mu$, and then takes $\sigma\to0$ at fixed renormalized parameters; since the new operator is irrelevant
in $d=4$, its contribution to IR observables is suppressed by $g_8(\mu)\sim\sigma^2\mu^4$ and vanishes smoothly.
In QM, the analogous statements are that (i) at fixed $\sigma$ the filter disappears in the continuum limit
$\Delta t\to0$ at fixed $t$, and (ii) at any fixed discretization the target unitary step is recovered directly
by sending $\sigma\to0$ on the relevant energy window.

Several concrete directions follow naturally from this operational viewpoint. On the QFT side, it would be useful
to work out explicit examples---e.g.\ compute how counterterms and renormalized observables approach their
$\sigma\to0$ limits in perturbation theory and in lattice discretizations where the local averaging can be
implemented directly. A closely related step is to formulate systematic lattice analogs: local averaging of
plaquette or link couplings produces gauge-invariant higher-dimensional operators, suggesting a controlled way to
construct symmetry-preserving improved or stabilized lattice actions. Finally, because the basic construction is a
completely positive Kraus map obtained from coherent control and postselection, it provides a controlled starting
point for exploring non-unitary extensions of QM and local QFT in which deviations from unitarity are constrained
by complete positivity, locality, and a well-defined unitary limit as $\sigma\to 0$.

\begin{acknowledgments}
This work was supported by the  European Union’s Horizon Europe research and innovation programme under grant agreement No. 101178170,  by the Israel Science Foundation under grant agreement No. 2208/24  and by the Israeli Council for Higher Education through ``QERNEL'' and the ``Interdisciplinary Center for the Theory of Quantum Computing''. 

\end{acknowledgments}

\appendix
\section{Proof of Proposition 1}
Let $M:=\max_j\|H_j\|$. Expand each exponential to second order:
\begin{align}
U_{\rm mix}(\Delta t)
&= \sum_j a_j \Big(I-iH_j\Delta t \notag\\
&\qquad\ - \tfrac12 H_j^2 \Delta t^2 + O(\Delta t^3 M^3)\Big),
\label{eq:appA1}\\
&= I-i\Big(\sum_j a_j H_j\Big)\Delta t \notag\\
&\quad - \tfrac12\Big(\sum_j a_j H_j^2\Big)\Delta t^2 + O(\Delta t^3 M^3),
\label{eq:appA2}
\end{align}
using $\sum_j a_j=1$. Meanwhile
\begin{equation}
e^{-i\bar H \Delta t}
= I-i\bar H \Delta t - \tfrac12 \bar H^{\,2}\Delta t^2 + O(\Delta t^3 M^3).
\label{eq:appA3}
\end{equation}

Subtracting \eqref{eq:appA3} from \eqref{eq:appA2} and taking the operator norm gives the stated bound.
(No Baker--Campbell--Hausdorff commutator terms arise here because $U_{\rm mix}$ is a \emph{linear combination}
rather than a product of exponentials; any noncommutativity is already contained in
$\bar H^{\,2}=(\sum_j a_j H_j)^2$.) See also product-formula results in \cite{Trotter1959}.

\section{Proof of Proposition~\ref{prop:continuum}}
Let $V_{\sigma,\Delta t} := e^{-iH\Delta t}\, e^{-\frac{1}{2}\sigma^2 \Delta t^2 H^2}$. Since both factors are bounded Borel functions of the same self-adjoint operator $H$, they \emph{commute} by the spectral theorem: $f(H)g(H)=(fg)(H)$. Hence
\begin{equation}
\begin{split}
\big(V_{\sigma,\Delta t}\big)^N
&= \Big(e^{-iH\Delta t}\Big)^N \Big(e^{-\frac{1}{2}\sigma^2 \Delta t^2 H^2}\Big)^N \\
&= e^{-iHt}\, e^{-\frac{1}{2}\sigma^2 t\,\Delta t\, H^2}.
\end{split}
\label{eq:appE1}
\end{equation}
\textbf{Fixed $\sigma$.} Let $\psi$ be any vector in the Hilbert space and write
$d\mu_\psi(E):=\mel{\psi}{P_H(dE)}{\psi}$ for the spectral measure of $H$ associated with $\psi$.
By the spectral theorem,
\begin{equation}
\begin{split}
\big\|\big(e^{-\frac{1}{2}\sigma^2 t\,\Delta t\, H^2}-\mathbbm{1}\big)\psi\big\|^2
&=\int_{\mathbb R}\Big|e^{-\frac{1}{2}\sigma^2 t\,\Delta t\,E^2}-1\Big|^2 \\
&\qquad\times d\mu_\psi(E).
\end{split}
\end{equation}
For each fixed $E$ the integrand tends to $0$ as $\Delta t\to 0$, and it is uniformly bounded by $4$
since $\big|e^{-a}-1\big|\le 2$ for all $a\ge 0$. Hence, by dominated convergence,
\[
\big\|\big(e^{-\frac{1}{2}\sigma^2 t\,\Delta t\, H^2}-\mathbbm{1}\big)\psi\big\|\xrightarrow[\Delta t\to 0]{}0
\qquad\text{for all }\psi,
\]
i.e.\ $e^{-\frac{1}{2}\sigma^2 t\,\Delta t\, H^2}\to\mathbbm{1}$ strongly. Using \eqref{eq:appE1}, it follows that
$\big(V_{\sigma,\Delta t}\big)^N\psi \to e^{-iHt}\psi$ for all $\psi$.
\smallskip

\noindent\textbf{Diffusive scaling.} If $\sigma^2=\kappa/\Delta t$ with $\kappa>0$ fixed, then \eqref{eq:appE1} becomes
\begin{equation}
\begin{split}
\big(V_{\sigma,\Delta t}\big)^N
&= e^{-iHt}\, e^{-\frac{\kappa t}{2} H^2} \\
&= \exp\!\Big(-iHt-\frac{\kappa t}{2}H^2\Big),
\end{split}
\label{eq:appE2}
\end{equation}
which holds \emph{exactly} for all $\Delta t>0$ because the two exponentials commute. \qed

\bibliographystyle{apsrev4-2}
\bibliography{interference_revtex_refs}

\end{document}